\documentclass[a4paper,twocolumn]{esapub} 
\usepackage{natbib}
\usepackage{epsfig}

\newcommand{\Msol}{M_{\sun}}
\newcommand{\fe}{\mbox{$^{60}$Fe}}
\newcommand{\co}{\mbox{$^{60}$Co}}
\newcommand{\nic}{\mbox{$^{60}$Ni}}
\newcommand{\al}{\mbox{$^{26}$\hspace{-0.2em}Al}}
\newcommand{\mg}{\mbox{$^{26}$Mg}}
\newcommand{\pcmq}{\mbox{cm$^{-2}$}}
\newcommand{\psec}{\mbox{s$^{-1}$}}
\newcommand{\funit}{\mbox{ph \pcmq \psec}}
\def\deg{\hbox{$^\circ$}}
\def\sun{\hbox{$\odot$}}

\title{Search for gamma-ray line emission from the radioactive decay of \fe\ 
       with SPI}
\author[1]{J. Kn\"odlseder}
\author[4,5]{E. Cisana}
\author[2]{R. Diehl}
\author[2]{G. Lichti}
\author[1]{M. Harris}
\author[1]{P. Jean}
\author[2]{K. Kretschmer}
\author[2]{A. von Kienlin}
\author[1]{J.-P. Roques}
\author[3]{S. Schanne}
\author[2]{V. Sch\"onfelder}
\author[1]{G. Vedrenne}
\author[1]{G. Weidenspointner}
\author[6]{C. Wunderer}
\affil[1]{Centre d'\'Etude Spatiale des Rayonnements, B.P. 4346, 31028 
          Toulouse Cedex 4, France (knodlseder@cesr.fr)}
\affil[2]{Max-Planck-Institut f\"ur extraterrestrische Physik, Postfach 1312, 
          85741 Garching, Germany}
\affil[3]{DSM/DAPNIA/Service d'Astrophysique, CEA Saclay, 91191 Gif-sur-Yvette, 
          France}
\affil[4]{IASF - CNR, via Bassini 15, 20133 Milan, Italy}
\affil[5]{Universit\`a di Pavia, Dipartimento di Fisica, via Bassi 6, 
          27100 Pavia, Italy}
\affil[6]{SSL, University of California Berkeley, CA 94720, USA}

\begin{document}

\keywords{gamma rays: observations, lines; supernovae; nucleosynthesis}

\maketitle

\begin{abstract}

The search for gamma-ray line emission from the radioactive decay of 
\fe\ figures among the prime scientific objectives of the INTEGRAL mission. 
\fe\ is believed to be primarily produced in core-collapse supernovae, although 
other sites, such as carbon deflagration supernovae or intermediate mass AGB 
stars have also been suggested. 
We present first results of our search for the 1173 and 1332 keV 
gamma-ray lines of \fe\ decay in the SPI data of the first mission year.
So far we can only report upper flux limits whose levels are determined 
by systematic uncertainties in the treatment of the instrumental 
background.
Once we understand the background uncertainties better, the SPI data 
should provide stringent constraints on \fe\ nucleosynthesis 
in our Galaxy.

\end{abstract}

\section{Introduction}
\label{sec:intro}

Searching for the gamma-ray lines at 1173.23 and 1332.49 keV that arise from 
the radioactive decay chain 
$\fe \to \co \to \nic$ figures among the prime objectives 
of the INTEGRAL mission. 
\fe\ is believed to be primarily produced in core-collapse supernovae, although 
other sites, such as carbon deflagration supernovae or intermediate mass AGB 
stars have also been suggested 
(see review in \citeauthor{knodlseder01}, 2001).

Theoretical estimates of \fe\ yields from core-collapse supernovae suffer from 
considerable uncertainties in nuclear reactions and in supernova 
explosion model details which affect these 
\citep{prantzos04}, hence observing the \fe\ gamma-ray lines will provide 
valuable constraints on supernova nucleosynthesis.
From comparisons of isotope abundance measurements which are co-produced 
with \fe, it has been estimated \citep{leising94} that a total amount 
of $0.9 \Msol$ should exist in the present Galaxy, leading to a flux of
$\sim 3 \times 10^{-5}$ \funit\ in each of the lines (corresponding to 
about 8\% of the flux in the 1809 keV line that arises from the radioactive 
decay chain $\al \to \mg$).
Recently, \citet{smith04b} has reported a first detection of the \fe\ lines by 
the solar observatory RHESSI, yet during this conference he revised the 
significance of the result to a marginal level of $2.6\sigma$ 
\citep{smith04a}.
By combining both lines, Smith derives a flux in each line of 
$(5.2 \pm 2.0) \times 10^{-5}$ \funit\ from the inner Galaxy, corresponding to 
about $10\%$ of the flux that RHESSI observes in the 1809 keV line.

In this paper we present our first results from the search for \fe\ decay 
emission in the data of the spectrometer SPI aboard INTEGRAL. 
For the moment we have no indications of significant \fe\ line emission, yet 
we are still facing considerable problems in the treatment of the instrumental 
background that have to be resolved before SPI can reach its actual sensitivity 
limit.

\section{Analysed data}

\begin{figure*}[!t]
  \center
  \epsfxsize=17cm \epsfclipon
  \epsfbox{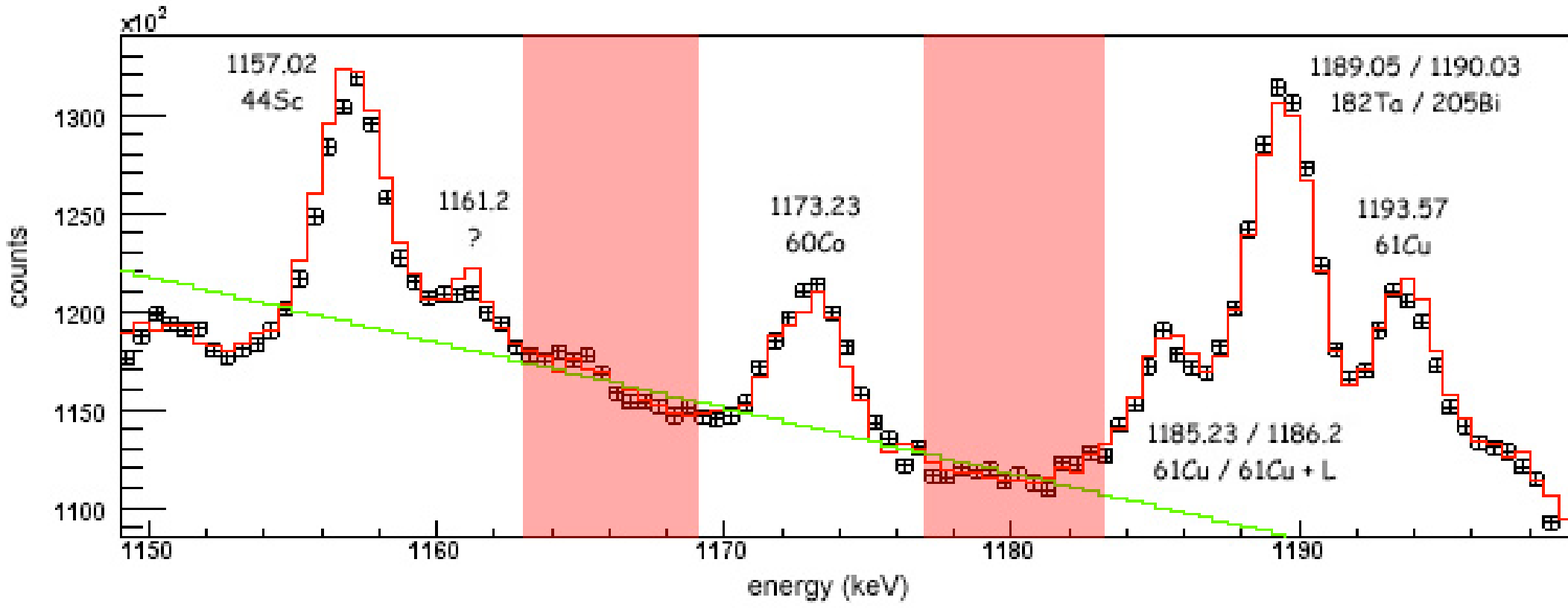}
  \epsfxsize=17cm \epsfclipon
  \epsfbox{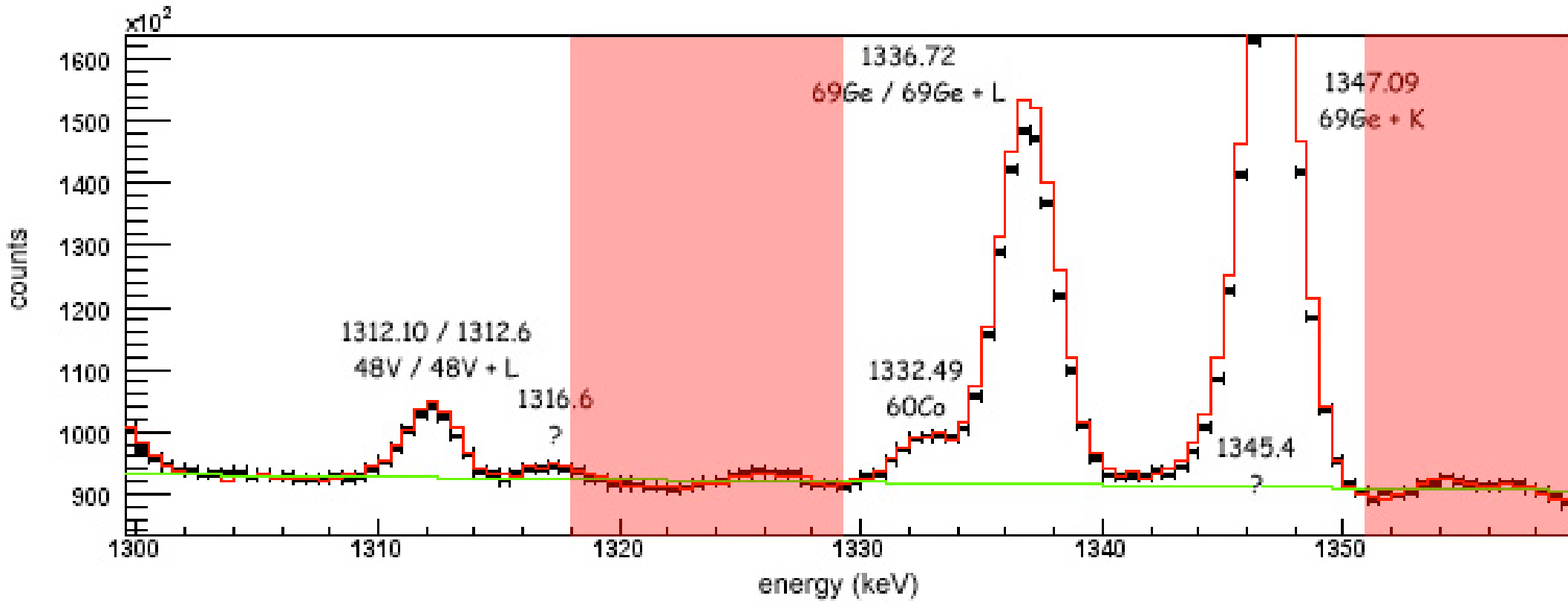}
  \caption{\label{fig:bgd}
  SPI instrumental background spectrum (crosses) in the area around the 
  \fe\ line energies together with the background model (histogram).
  The summed and degradation corrected data for GCDE1 and 2 are shown.
  The grey areas (red in the colour version of the paper) indicate the 
  energy bands that have been used for modelling 
  the continuum emission underlying the line (indicated by the light 
  line). 
  Instrumental background lines have been labelled by their source isotopes, 
  where identified. 
  Note that we only attempt to model the \co\ background lines, 
  while other background lines which obey different time variabilities
  are not described satisfactorily.
  }
\end{figure*}

The data that we analysed in this work consist of the INTEGRAL first year's
spring and autumn Galactic Centre Deep Exposures (GCDE1 and 2), with 1.8 
and 2.0 Ms of exposure, respectively, covering galactic longitudes 
$l = \pm 30\deg$ and latitudes $b = \pm 20\deg$. 
GCDE1 includes INTEGRAL orbits 47-66, GCDE2 includes orbits 97-123. 
Only the analysis of single-detector event data is presented in this 
paper. 
The analysis of multiple-detector event data has been started, yet due 
to issues concerning the background modelling we have so far focused our 
attention on the single-detector event data. 
Energy calibration has been performed orbit-by-orbit, resulting in a relative 
(orbit-to-orbit) calibration precision of $\sim0.01$ keV and an absolute 
accuracy of $\sim0.05$ keV \citep{lonjou04}.

\section{Background modelling}

Background modelling is performed in our usual approach, where a two 
component model aims to describe both the instrumental line and continuum 
components. 
At the rest energies of the astrophysical \fe\ lines (1173.23 and 1332.49 keV) 
the SPI background (cf.~Fig.~\ref{fig:bgd}) shows two instrumental line features
that have been identified as activation lines of the radio-isotope \co\
(i.e. the same isotope that we aim also to observe in space; 
\citeauthor{weidenspointner03} 2003).

For the line component, the spectral shape and the relative counting rates 
in each detector have been estimated using empty field observations, assuming 
that both do not vary in time (in fact, detector degradation leads to a 
modification of the instrumental line shapes that is corrected for 
after background modelling; \citeauthor{knodlseder04} 2004).
A total empty field exposure of 4.1 Ms has been collected by gathering all 
data for which SPI pointed at galactic latitudes $|b| > 20\deg$ (covering 
INTEGRAL orbits 14-106). 
To extract the line component, a linear continuum background has been 
subtracted from the spectra, estimated from adjacent energy bands 
(1163-1169 \& 1177-1183 keV for the continuum under the 1173 keV line and 
1318-1329 \& 1351-1360 keV for the continuum under the 1332 keV line). 

\begin{figure}[!b]
  \center
  \epsfxsize=8cm \epsfclipon
  \epsfbox{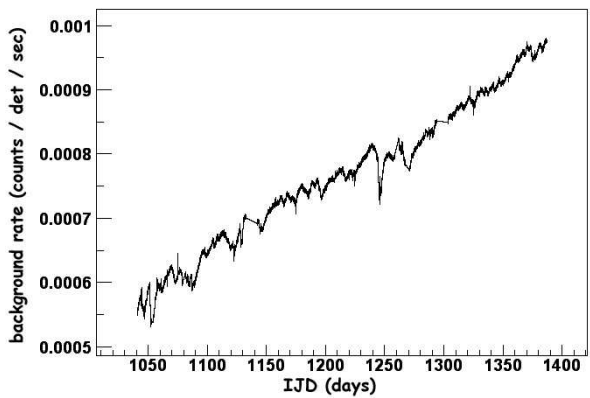}
  \caption{\label{fig:template}
  Background variation template of the instrumental background line at
  1173.23 keV.
  }
\end{figure}

\begin{figure*}[!t]
  \center
  \epsfxsize=17cm \epsfclipon
  \epsfbox{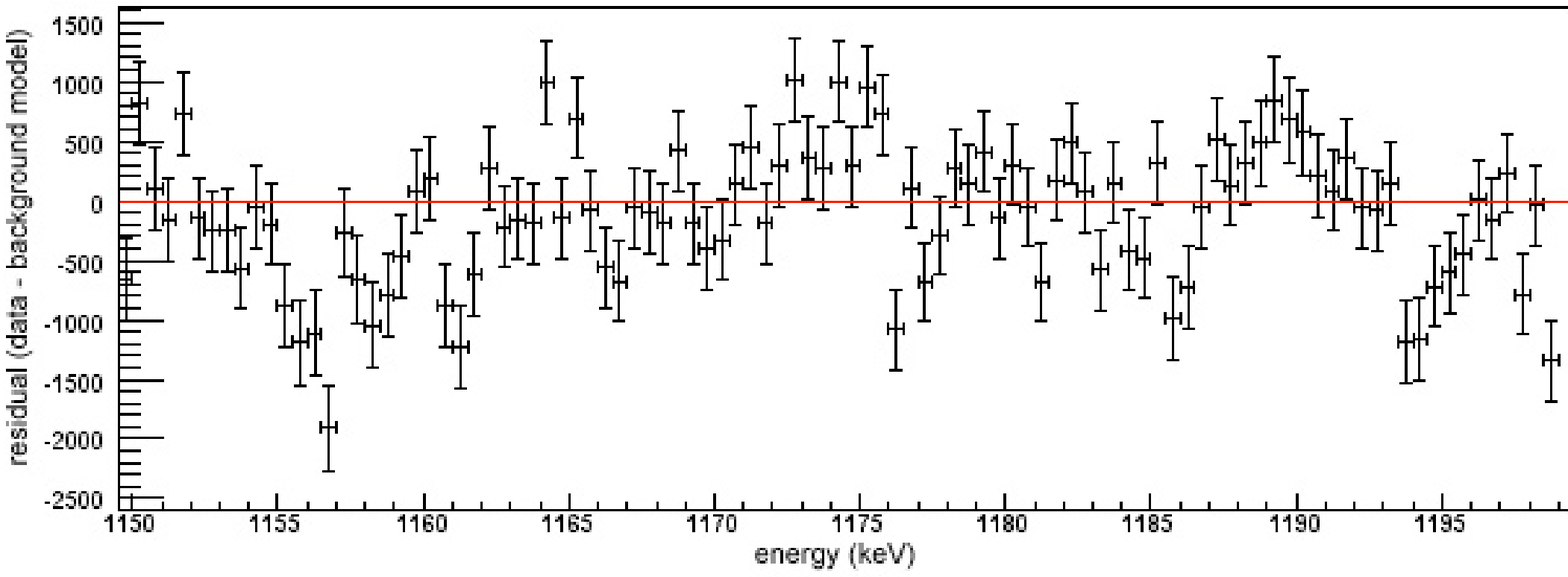}
  \center
  \epsfxsize=17cm \epsfclipon
  \epsfbox{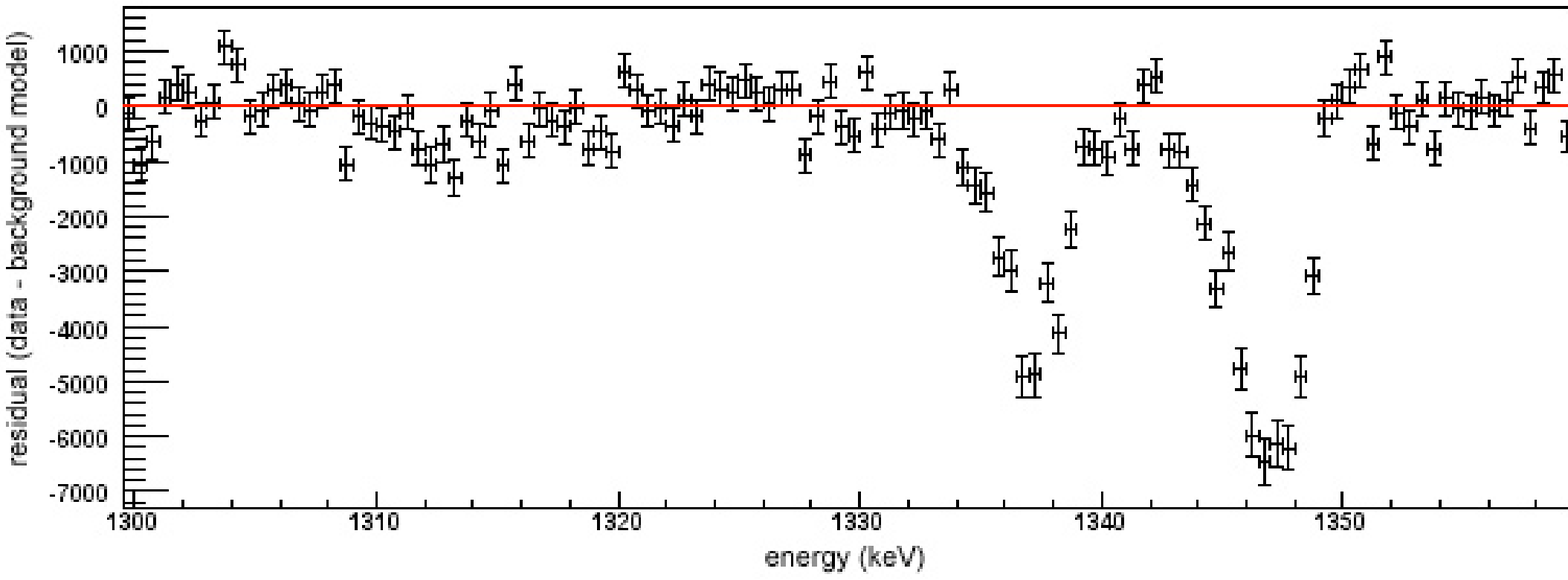}
  \caption{\label{fig:residuals}
  Residual SPI energy spectrum after subtraction of the background 
  model (cf.~Fig.~\ref{fig:bgd}).
  }
\end{figure*}

The time variability of the line component has been modelled by what we 
call a background variation template. 
This template is a combination of background activity tracers that have 
been adjusted to explain the line background variation in the empty field 
observations.
The idea behind this approach is that an instrumental background line 
has in general $N$ physical contributions which may be prompt 
(instantaneous) or may eventually come from decay chains with various decay 
constants $\tau$.
Thus, in general, the counting rate of the instrumental background line 
component can be written as
\begin{equation}
 r(t) = \sum_{i=1}^N a_{i} \frac{1}{\tau_{i}} 
        \int_{0}^{t} r_{i}(t') e^{-(t-t')/\tau_{i}} dt'
\end{equation}
where the index $i$ runs over the $N$ contributions to the background 
line, $r_{i}(t)$ is the counting rate of the activity tracer used to 
model the contribution $i$, $\tau_{i}$ is the decay constant of 
contribution $i$, and $a_{i}$ is the relative amplitude of contribution $i$.
The time integral is taken from the start of the INTEGRAL mission up 
to the date of interest $t$.

\begin{figure*}[!t]
  \center
  \epsfxsize=8.4cm \epsfclipon
  \epsfbox{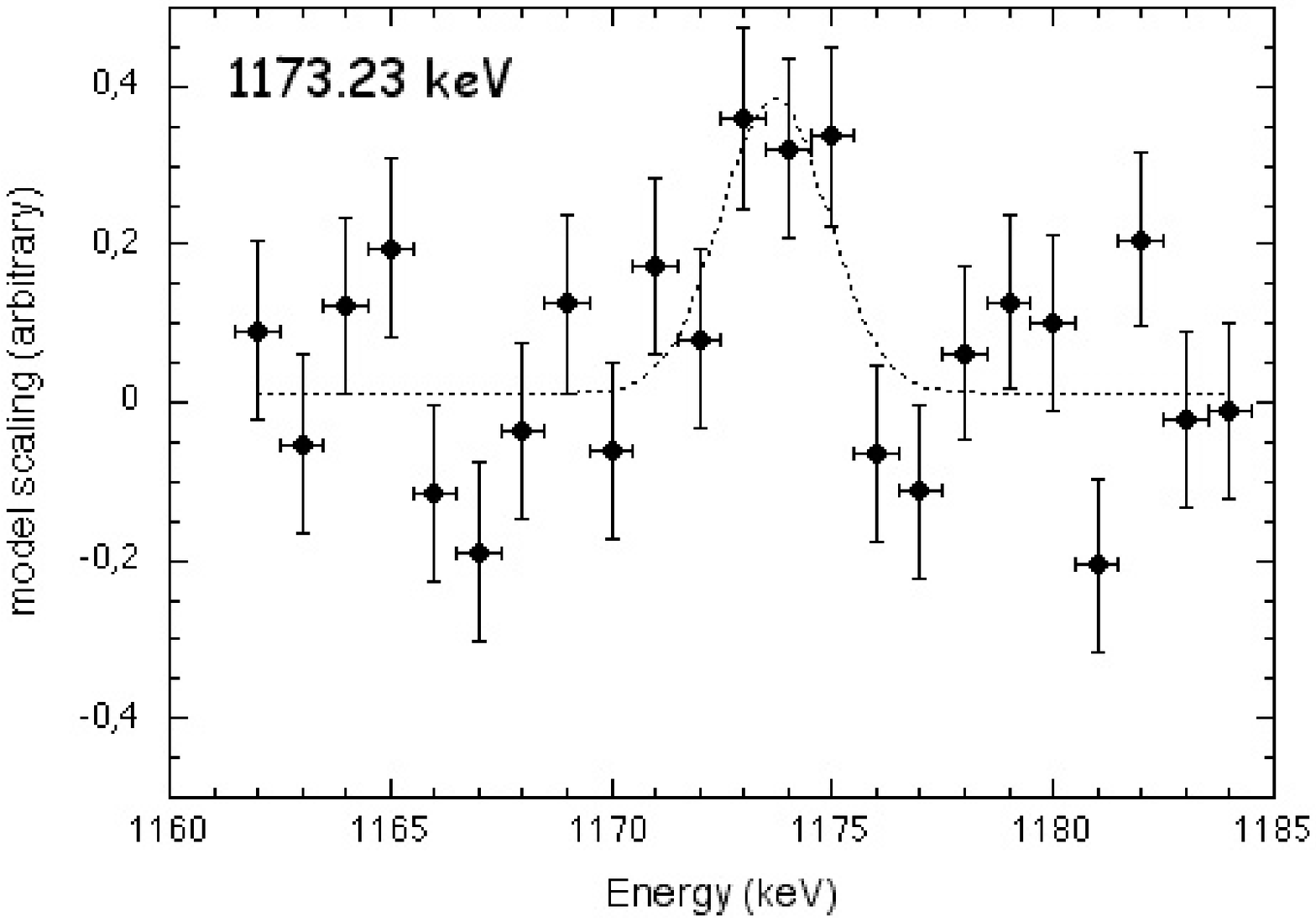}
  \hfill
  \epsfxsize=8.4cm \epsfclipon
  \epsfbox{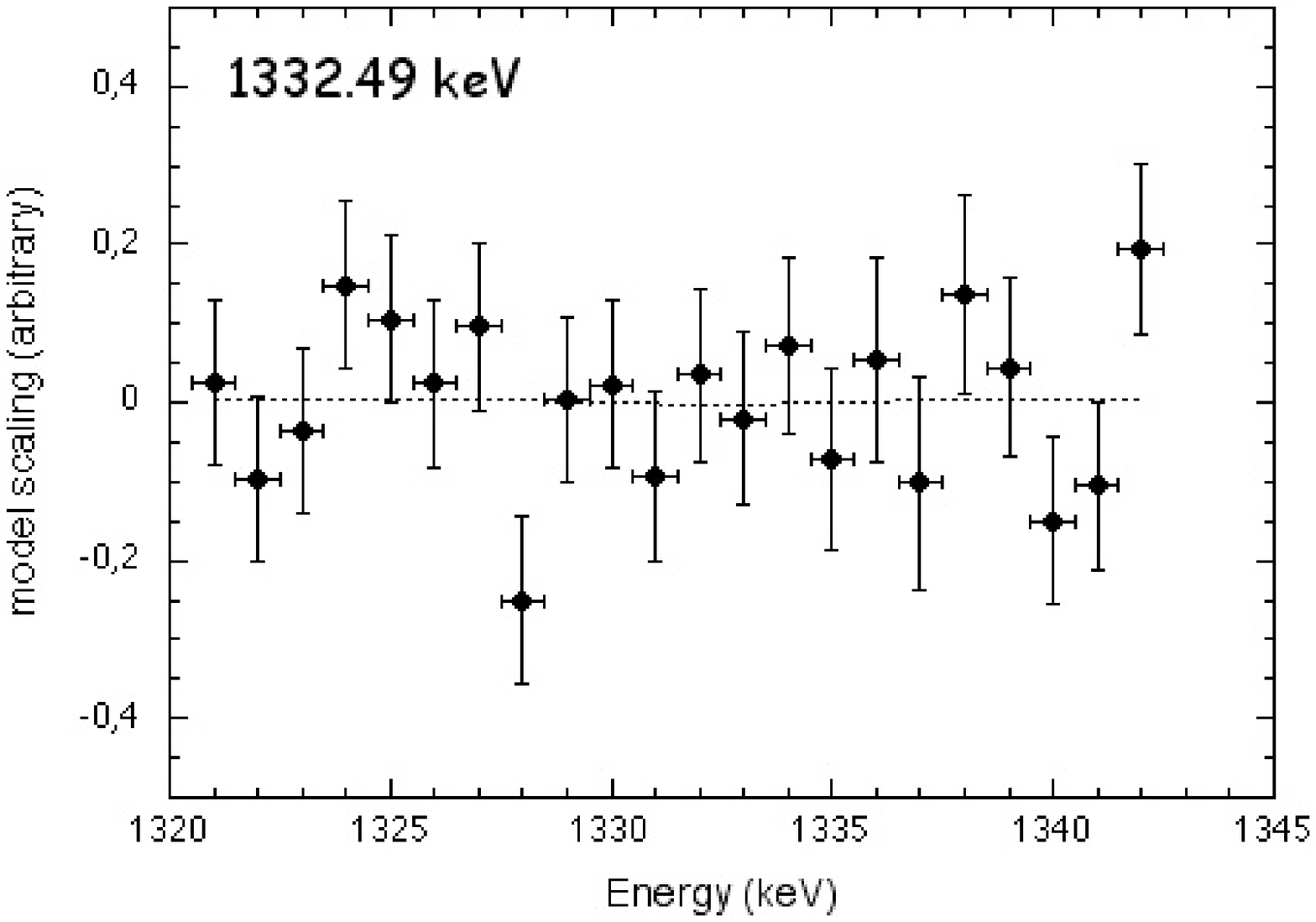}
  \caption{\label{fig:spectra}
  SPI spectrum of the galactic centre region (GCDE1 and 2), assuming an \fe\ 
  line intensity distribution that follows the Dirbe 240 $\mu$m infrared 
  emission as a tracer of the massive star population.
  The ordinate gives the \fe\ line flux in units of $10^{-4}$ \funit\ 
  keV$^{-1}$ for the central steradian of our Galaxy.
  The residual of the instrumental background line at 1337 keV from
  $^{69}$Ge has been modelled by fitting a Gaussian 
  function to the spectrum and by subtracting this function from the
  data points.
  }
\end{figure*}

For the \fe\ analysis, we tried different combinations of activity 
tracers such as the germanium saturated event rate as a tracer of the 
high-energy particle irradiation of the SPI telescope or the plastic 
scintillator event rate as a tracer of the low-energy particle 
irradiation.
As example, we show in Fig.~\ref{fig:template} a template made of two 
contributions, which consist of 
a prompt background component (i.e. $\tau = 0$), modelled using the germanium 
saturated event rate, and a \co\ activation component, modelled using the 
germanium saturated event rate convolved with an exponential law with a time 
constant of $\tau = 7.6$ yr.
The relative amplitudes $a_{i}$ of each contribution have been 
adjusted using the empty field data.
Replacing the germanium saturated rate by the plastic scintillator event 
rate or adding further components does not alter our background model 
significantly.
Therefore we only show results in this paper that have been obtained 
by using the template of Fig.~\ref{fig:template}.

The continuum background has been estimated by adjusting linear functions 
to the data in energy bands adjacent to the line (the same energy bands that 
have been used for the line component extraction in the empty field data). 
In principle, this adjustment should be done for each detector and pointing, 
yet the small number of counts in the dataspace requires an adaptive rebinning 
(in the time axis), requesting a minimum number of 1000 events per adjacent 
energy bin. 
This rebinning has been found to be crucial in reducing systematic uncertainties 
in the line shape analysis.
As long as the number of events used to determine the continuum 
level is sufficiently large ($>1000$), the precise smoothing 
parameters have little impact on the result.

Figure \ref{fig:residuals} shows the background subtracted spectrum 
for the summed GCDE1 and 2 data.
As we attempted only to model the instrumental \co\ lines, strong 
residuals remain at the energies of instrumental background lines that 
have time variability different from that of \co.
In particular, the prompt background lines of $^{69}$Ge at 1337 and
1347 keV show up as strong negative residual lines in the spectrum.
Note that also at 1173 keV residual counts remain, while at 1332 keV 
the spectrum appears well-represented (insignificant residuals).

\section{Spectral analysis}

Straight background spectra subtraction does not make use of the coded-mask 
and field-of-view properties of SPI; on the other hand, image deconvolution 
is feasible only for strong signals. 
As we expect a weak or no signal, we derive spectra by fitting the 
amplitude of a given skymap of (expected) intensity distribution, per 1 keV 
wide energy bin. 
As skymap we used the Dirbe 240 $\mu$m intensity distribution, since this map 
has been shown to provide a reasonable tracer of galactic 1809 keV emission, 
and hence is probably also a reasonable first-order tracer of the \fe\ 
distribution \citep{knodlseder99}. 
Fig.~\ref{fig:spectra} shows the resulting spectra. 

As in the residual spectrum (cf.~Fig.~\ref{fig:residuals}), a significant 
line feature appears at 1173 keV. 
The formal flux in this feature amounts to 
$(9.3 \pm 2.6) \times 10^{-5}$ \funit, which corresponds to $\sim30\%$ of the 1809 keV 
line flux. 
This is three times as much as the flux reported by RHESSI 
\citep{smith04a}) and also exceeds previously reported upper flux limits 
\citep{diehl98}. 
At 1332 keV no line feature appears.
The formal 1332 keV line flux amounts to 
$(-0.2 \pm 2.4) \times 10^{-5}$ \funit.

\section{Discussion and conclusions}

If the feature at 1173 keV were a real celestial signal, we would have 
expected to find the same flux values for both lines.
In addition, the formal flux in this features exceeds 
the expectations based on measurements by other instruments.
Therefore we conclude that the spurious line residual at 1173 keV, 
inconsistent with \fe\ due to the absence of a corresponding signal at 
1332 keV, should be used to determine our systematic uncertainty at this 
point. 
From this, we derive an upper flux limit in each of the \fe\ lines of 
about $10^{-4}$ \funit. 
In comparison, the statistical uncertainty of $\sim 2.5 \times 10^{-5}$ 
\funit\ is small.

Our current data analysis efforts are concentrated on reducing the 
systematic uncertainties and improving the subtraction of the instrumental 
background lines.
Once the systematic uncertainties have been reduced to a reasonable 
level (i.e. below the statistical uncertainties of the data) we can 
expect $2\sigma$ upper flux limits of the order of $5 \times 10^{-5}$ \funit\
in each of the \fe\ lines for the first year of GCDE data.
Combining both lines would reduce the upper flux limit by roughly
$\sqrt{2}$, leading to a level of $3.5 \times 10^{-5}$ \funit.
Exploiting also the multiple-event data (which obey about the same 
signal-to-noise ratio as the single-event data) will give another 
factor of $\sqrt{2}$, providing an upper flux limit of 
$2.5 \times 10^{-5}$ \funit.
This is about $6-8\%$ of the 1809 keV line flux for each of the 2 
lines
(dependent on which 
flux level is assumed for the 1809 keV line intensity) and would 
present the most stringent upper flux limits available today.
Adding more data to our still rather small dataset (only 3.8 Ms
of exposure time have been exploited so far, yet much more data have 
already been collected along the galactic plane by INTEGRAL) will further 
reduce the sensitivity limit.

On the other hand, as stated in section \ref{sec:intro}, cosmic 
abundance arguments suggest a \fe\ line flux of $\sim 3 \times 10^{-5}$ 
\funit\ in each of the lines.
This is comparable to the $2\sigma$ upper limit that is achievable by 
SPI using the first year's GCDE data and is close to the value 
suggested by RHESSI.
Adding all INTEGRAL exposure of the central galactic regions from the 
first mission year will probably bring the expected \fe\ line 
intensity near to a firm detection.
Thus, it is highly likely that the spectrometer SPI on INTEGRAL will provide 
in the near future serious constraints on \fe\ nucleosynthesis in our 
Galaxy.


\end{document}